\documentclass[referee]{aa}
\usepackage{psfig}
\begin{document}

   \thesaurus{06     
              (02.01.2;  
               08.02.1;  
               08.09.2 (GRO\,J1655$-$40);  
               13.25.5)}  

   \title{Turmoil on the accretion disk of GRO\,J1655$-$40}

   \author{E.~\mbox{Kuulkers}
           \inst{1,2}
	   \and
	   J.J.M.~in 't Zand
	   \inst{1}
           \and
           R.~Cornelisse
           \inst{1,2}
           \and
           J.~Heise
           \inst{1}
           \and
           A.K.H.~Kong
           \inst{3}
	   \and
           P.A.~Charles
           \inst{3,4}
           \and
           A.~Bazzano
           \inst{5}
           \and
           M.~Cocchi
           \inst{5}
           \and
           L.~Natalucci
           \inst{5}
           \and
           P.~Ubertini
           \inst{5}
          }

   \offprints{Erik \mbox{Kuulkers}}

   \institute{
	      Space Research Organization Netherlands,
	      Sorbonnelaan 2, NL-3584 CA Utrecht, The Netherlands\\
	      email (EK): E.\mbox{Kuulkers}@sron.nl
              \and
              Astronomical Institute, Utrecht University, 
              P.O.\ Box 80000, NL-3508 TA Utrecht, The Netherlands
              \and
              Department of Astrophysics, Nuclear \& Astrophysics Laboratory,
              Keble Road, Oxford OX1 3RH, United Kingdom
              \and
              Department of Physics \&\ Astronomy, University of Southampton, 
              Southampton, SO17 1BJ, UK 
              \and
              Istituto di Astrofisica Spaziale (CNR), Area Ricerca Roma Tor Vergata,
              Via del Fosso del Cavaliere, 
              I-00133 Roma, Italy
             }

   \date{Received 10 February 2000; accepted --}

   \titlerunning{Turmoil in GRO\,J1655$-$40}

   \maketitle

   \begin{abstract}

During the 1996/1997 outburst of the X-ray transient GRO\,J1655$-$40
the {\em Wide Field Cameras} onboard BeppoSAX 
observed the system with nearly continuous coverage during some stages 
of this outburst. 
On two occasions we find clear evidence for frequent
dipping behaviour, characterized by several absorption dips 
occurring within one binary orbit.
However, during preceding and following binary orbits there is no 
sign of such dips at the same range of orbital phases.
We also collected light curves of absorption dips as seen with
other X-ray instruments.
Although the 65 dips that were collected occur preferably 
between orbital phases 0.68 and 0.92, there is evidence that
the morphology of the material causing 
the dips is changing over just one binary orbit.
The short duration ($\sim$minutes) and phasing of the dips
are discussed in the framework of models where the stream from the companion star
encounters a hot outer rim on the accretion disk.

      \keywords{accretion, accretion disks --- binaries: close ---
                stars: individual (GRO\,J1655$-$40) --- X-rays: stars
               }
   \end{abstract}

\section{Introduction}

The black-hole candidate and soft X-ray transient GRO\,J1655$-$40 
(X-ray Nova Sco 1994) was discovered on 1994 July 27 by the 
Burst and Transient Source Experiment (BATSE) onboard the 
{\em Compton Gamma Ray Observatory} (Zhang et al.\ 1994, Harmon et al.\ 1995). 
Soon thereafter the optical 
counterpart was found (Bailyn et al.\ 1995a). It has since shown irregular outburst activity
(e.g.\ Zhang et al.\ 1997). After the 1995 July/August hard X-ray outburst the source
settled back to quiescence. However, on 1996 April 25 it became active again 
(Remillard et al.\ 1996, Levine et al.\ 1996, Orosz et al.\ 1997) for a period 
which lasted $\sim$16~months (see e.g.\ Remillard et al.\ 1999).
GRO\,J1655$-$40 has an orbital period of 2.62~days, has  
dynamically been shown to contain a black hole with a mass of 
$\sim$7\,M$_{\odot}$, and is viewed at an inclination of $\sim$70$^{\circ}$
(Bailyn et al.\ 1995b, Orosz \&\ Bailyn 1997, van der Hooft et al.\ 1998, Shahbaz et al.\ 1999).
The object has become notorious for being one of the systems showing
relativistic radio jets, during its first outburst in 1994 (Tingay et al.\ 1995; 
Hjellming \&\ Rupen 1995). 

Deep short dips in the X-ray light curves of GRO\,J1655$-$40 were discovered during its 1996/1997 outburst 
These dips appeared periodically between orbital 
phases $\sim$0.7 and $\sim$0.85 and
were shown to be due to a medium absorbing the radiation coming from the inner
parts of the accretion disk (\mbox{Kuulkers} et al.\ 1998a).
Such ``orbital'' dips are known to occur also in various other
low-mass and high-mass X-ray binaries (see Parmar \&\ White 1988; 
Marshall et al.\ 1993; Saraswat et al.\ 1996; and references therein).
They probe the interaction between the stream from the companion 
and the accretion disk, as well as the X-ray emitting region itself
(see e.g.\ \mbox{Kuulkers} et al.\ 1998a; Tomsick et al.\ 1998).

We report here on observations made with the {\it Wide Field Cameras} (WFC) 
onboard the {\em Beppo Satellite per Astronomia X} (BeppoSAX) 
observatory, obtained during various stages of the last outburst in 1996/1997 of GRO\,J1655$-$40. 
Due to the relatively long orbital period of GRO\,J1655$-$40, the WFC is ideally suited
for monitoring such a system at all orbital phases almost continuously, which
is not possible or undertaken with other X-ray instruments carrying out
short programs of typically less than a day or with all-sky monitors
viewing sources typically up to ten times a day for a short 
period of order minutes. We summarize all dip events seen during the 1996/1997 outburst 
with the WFC and other instruments and discuss their origin.

\section{Observations and Analysis}

The Wide Field Camera (WFC) instrument (Jager et al.\ 1997) consists of two 
coded aperture cameras that point in opposite directions and perpendicular 
to the Narrow-Field Instruments on the same BeppoSAX satellite (Boella et
al.\ 1997). The field of view of each camera is 40 by 40 square degrees
full-width to zero response with an angular resolution about 5 arcminute
in each direction. The detectors are Xenon-filled multi-wire proportional counters
with band passes of 2 to 28\,keV. The sensitivity in the 2--10\,keV band is about 10\,mCrab in 
10$^4$\,s. The imaging capability and sensitivity
allow an accurate monitoring of complex sky regions, like the
Galactic bulge. The WFCs are carrying out a program of
monitoring observations of the field around the Galactic Center.
The purpose is to detect X-ray
transient activity, particularly from low-mass X-ray binaries (LMXBs)
whose Galactic population exhibits a strong concentration in this field,
and to monitor the behaviour of persistently bright X-ray sources
(see e.g.\ Heise 1998). This program consists of campaigns during the 
spring and autumn of each year. Each campaign lasts about two months and typically
comprises weekly observations.
The favorable position of GRO\,J1655$-$40 
($l=344\fdg 98$, $b=+2\fdg 46$ or 
${\rm R.A.} = 16^h 54^m 00^s, {\rm Dec.} = -39\degr 50\arcmin 45\arcsec$, J2000.0) 
within the vicinity of the Galactic Center implies therefore a relatively large
coverage of the source during its 1996/1997 outburst.

In this paper we report on part of the X-ray monitoring program
during 1996 August-October and 1997 March-April and August, as well as 
serendipitous observations close to the Galactic Center (see Table 1),
yielding a total of 1.37\,Msec of good data. 
These observations were either performed with WFC unit 1 or 2.
For the light curves we used as intensity measure the count rate
per cm$^2$ in the 2--8\,keV band. All measurements are
corrected for dead time. 

\begin{table}
\caption{BeppoSAX WFC observation log of GRO\,J1655$-$40}
\begin{tabular}{cc}
\hline
Date & JD$-$2450000 \\
\hline
1996 Aug 20--30 & 315--325\\
1996 Sep  2--4; 12--13; 15--18; 22-25 & 328--330; 338--339; 341--344; 348--351\\
1996 Oct 5; 10--13 & 361; 366--369 \\
1997 Mar  7--8; 13--20; 23--25; 29--31 & 514--515; 520--527; 530--532; 536--538 \\
1997 Apr 1; 7; 11; 13--15 & 539; 546; 548; 550--553 \\
1997 Aug 12--13 & 672--673 \\
\hline
\end{tabular}
\end{table}

The All-Sky Monitor (ASM; Levine et al.\ 1996) onboard the {\em Rossi X-ray Timing Explorer} 
(RXTE) scans the X-ray sky in series of 90\,sec dwells in three energy bands between
2 and 12\,keV. A given source is observed typically 5 to 10 times a day. For our purpose
we used the ASM quick-look results in the 2--12\,keV band as provided by the RXTE/ASM team at 
the Massachusetts Institute of Technology (MIT).

Numerous pointed target-of-opportunity observations were obtained during the
1996/1997 outburst of GRO\,J1655$-$40 with the
Proportional Counter Array (PCA: 2--60\,keV; Bradt et al.\ 1993) onboard RXTE
(Sobczak et al.\ 1999; Remillard et al.\ 1999). 
Dips in the light curves were seen in observations taken on 1996 June 20
(Hynes et al.\ 2000, in preparation), 1997 January 5
(Remillard et al.\ 1999) and 1997 February 26 (\mbox{Kuulkers} et al.\ 1998a,b).
We used the data collected with a time resolution
of 0.125\,sec in the total 2--60\,keV energy band from the so-called `Standard~1' mode.

The High Resolution Imager (HRI: 0.1--2.4\,keV; e.g.\ Zombeck et al.\ 1995) onboard
the {\em R\"ontgen Satellite} (ROSAT) 
observed GRO\,J1655$-$40 various times in the periods 1996 August 28 to September 24 
and 1997 March 10 to 13. During 
the second epoch deep short drops in intensity were found on March 12, similar
to those seen with the RXTE/PCA. For the light curve we used all counts registered
by the HRI; we did not correct for background, vignetting and dead-time,
since the source is bright and since we are mainly interested in the dip light curves
and times of dip occurrence.

The {\it Advanced Satellite for Cosmology and Astrophysics} (ASCA) 
satellite (Tanaka et al.\ 1994) is equipped with 
two Solid State Imaging Spectrometers (SIS; 0.4--10\,keV) and two Gas Imaging 
Spectrometers (GIS; 0.7--10\,keV). 
For our purposes we only used the data taken by the GIS instruments 
since they provide a higher time resolution compared to the SIS instruments. 
ASCA observed GRO\,J1655$-$40 during its 1996/1997 outburst 
on 1997 February 26--28. The GIS instruments were set to PH mode with timing resolution
of 62.5\,ms in high bit rate and 0.5\,s in medium bit rate. Standard data
screening was employed\footnote{See 
the {\it ASCA} Data Reduction Guide, Version 2.0\\
(http://heasarc.gsfc.nasa.gov/docs/asca/abc/abc.html).}.
Data taken at a geomagnetic cut-off rigidity lower than 4\,GeV, at an elevation
angle less than $5^{\circ}$ from the Earth limb and during passage through the
South Atlantic Anomaly were rejected. After filtering, the total net exposure
time of each GIS was 33.7\,ks. We extracted the GIS light curves from a
circular region of radius $6'$ centered at the position of the source. 

\section{Results}

\subsection{The quest for X-ray dips}

In the upper panel of Fig.~1 we show the outburst light curve of 
GRO\,J1655$-$40 as obtained with the ASM. The outburst
seems to consist of two stages separated by JD\,2450450 (1997 January 1).
The first stage shows strong flaring behaviour, whereas the second stage
shows a smoother light curve. In the lower
panel of Fig.~1 we show the observations as obtained with the WFC. 
Note that the last set of WFC observations were taken just before the source 
returned to quiescence (see also Sobczak et al.\ 1999, Remillard et al.\ 1999).
The deep drops clearly seen in the ASM light curve are due to absorption
dips, previously identified by \mbox{Kuulkers} et al.\ (1998a).
In the WFC light curve such dipping behaviour is clearly present 
at the end of the second period of WFC observations. 

\begin{figure}
\centerline{
\psfig{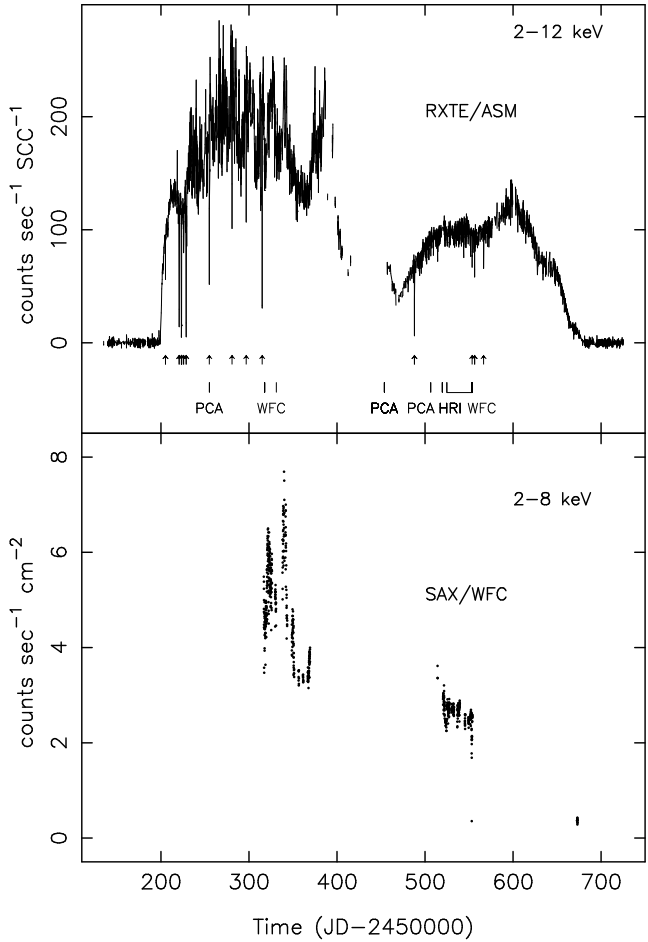}
}
\caption{
Upper panel: RXTE/ASM light curve of the 1996/1997 outburst of GRO\,J1655$-$40.
Shown are the 90-sec measurements; data points lying less than $\sim$2\,d apart are connected by a line 
to guide the eye.
Clearly, deep drops can be seen which are due to the absorption dips; they are indicated
by arrows at the lower part of this panel. Also at the lower part of this panel 
are indicated the times at which absorption dips were seen with other instruments (see Table 2).
Lower panel: BeppoSAX/WFC coverage of the 1996/1997 outburst. Shown are averages over one
satellite orbit. JD\,2450200 corresponds to 1996 April 26, 12:00 (UT).
}
\end{figure}

We constructed WFC light curves with a time resolution of 5\,sec, and searched for
clear drops in intensity. 
On several occasions we found clear dipping behaviour in the X-ray light curves; 
they all appeared between orbital phases 0.68 and 0.89. 
During the observations on 1996 August 22 and 1997 April 14
various dips appeared; their detailed light curves
are given in Figs.~2 (left panel) and 3 (right panel). On other occasions 
only single dips were found (see Table 2).
It is clear that the dips are very deep, i.e.\ down to $\sim$1\%\ of the out-of-dip
intensity, as was already noticed for other dips from
RXTE/PCA and ASM observations discussed by \mbox{Kuulkers} et al.\ (1998a).
Their durations range from $\sim$15\,sec to $\sim$3.5\,min.
The profiles of the dips are reminiscent
of those of the dips seen previously with the RXTE PCA (\mbox{Kuulkers} et al.\ 1998a).
Particularly interesting is the light curve of 1997 April 14 (Fig.~3, right panel), 
during which continuing dip activity occurred for a total time span of $\sim$9.2\,hr,
i.e.\ between orbital phases 0.68--0.79 of binary cycle\footnote{Binary cycle is defined as the 
number of binary orbits since JD\,2449838.4198 (van der Hooft et al.\ 1998).} 272. 

\begin{figure}
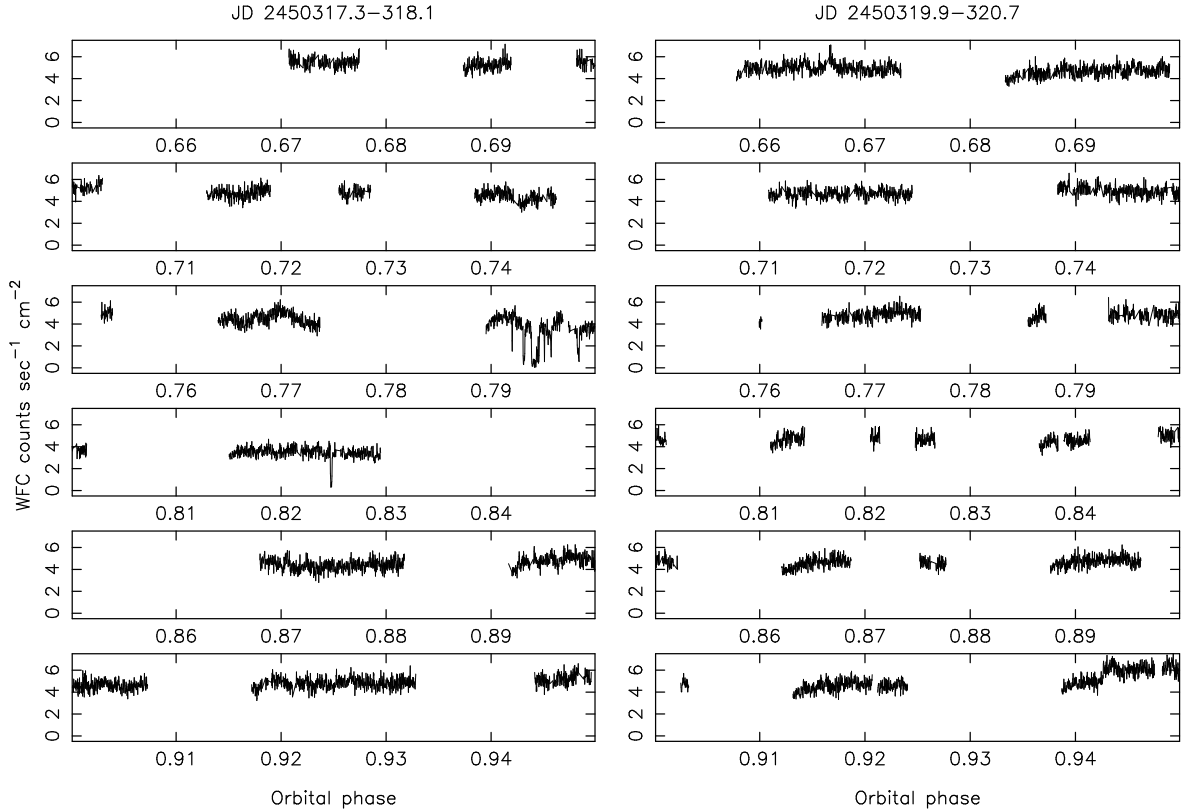

\centerline{
\hspace{2cm}
\psfig{figure=1655_wfc_fig2a.ps,bbllx=20pt,bblly=58pt,bburx=539pt,bbury=766pt,width=8cm}
\hspace{.0cm}
\psfig{figure=1655_wfc_fig2b.ps,bbllx=50pt,bblly=58pt,bburx=539pt,bbury=766pt,width=7.555cm,clip=yes}
}
\caption{
WFC light curves of GRO\,J1655$-$40 from two consecutive binary orbits at 
$\phi_{\rm orb}$=0.65--0.95 near the start of the WFC observations 
(JD\,2450317 [left] and JD\,2450320 [right] correspond to 1996 August 21 and 
24, respectively, at 12:00 UT). 
The time resolution is 5\,sec.
In the left panel dips can be seen near $\phi_{\rm orb}$=0.79--0.80, whereas
one binary orbit later (right panel) they are absent at the same orbital phase 
range.
}
\end{figure}

\begin{figure}
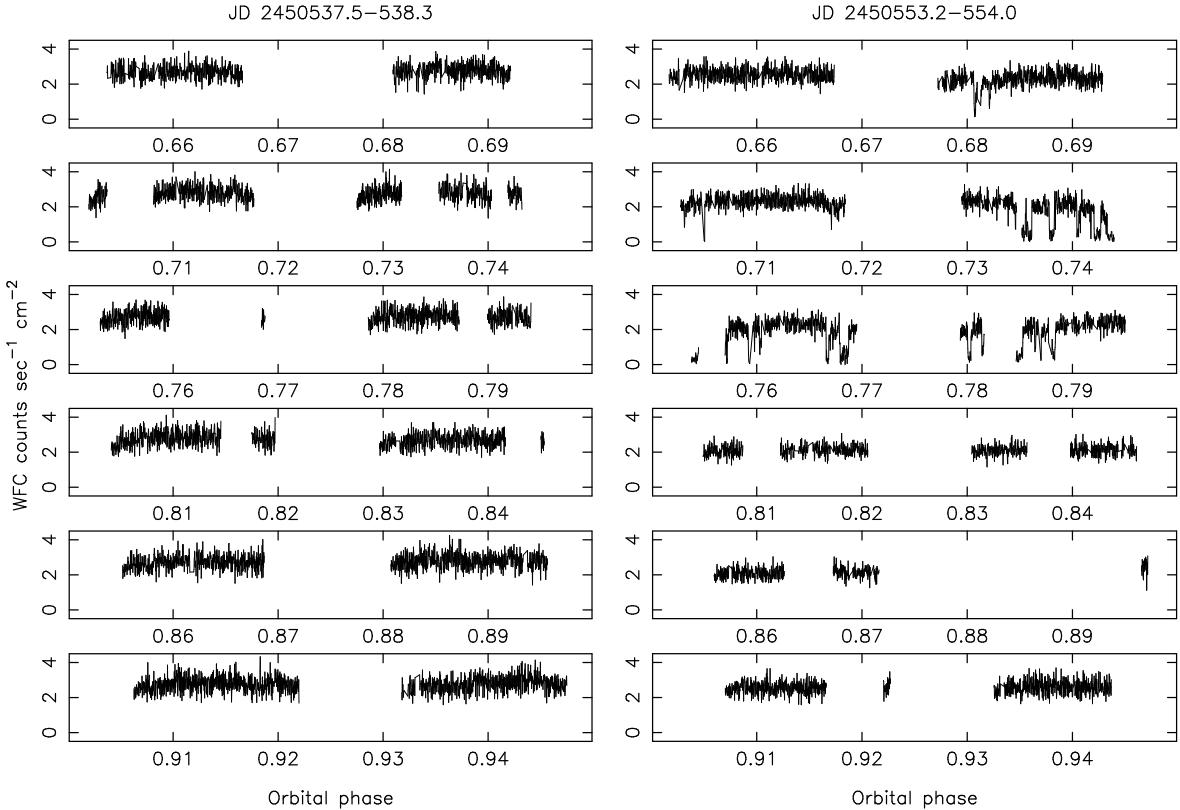

\centerline{
\hspace{2cm}
\psfig{figure=1655_wfc_fig3a.ps,bbllx=20pt,bblly=58pt,bburx=539pt,bbury=766pt,width=8cm}
\hspace{.0cm}
\psfig{figure=1655_wfc_fig3b.ps,bbllx=50pt,bblly=58pt,bburx=539pt,bbury=766pt,width=7.555cm,clip=yes}
}
\caption{
WFC light curves of GRO\,J1655$-$40 from $\phi_{\rm orb}$=0.65--0.95, which are taken
6 binary orbits apart near the end of the second part of WFC observations
(JD\,2450538 [left] and JD\,2450320 [right] correspond to 1997 March 30 and 
April 14, respectively, at 12:00 UT). 
The time resolution is 5\,sec. Note that the y-axis scale is different from that in Fig.~2.
In the right panel dips can be seen from $\phi_{\rm orb}$=0.68--0.79, whereas
6 binary orbits earlier (left panel) they are absent at the same orbital phase 
range. 
}
\end{figure}

\begin{figure}
\centerline{
\psfig{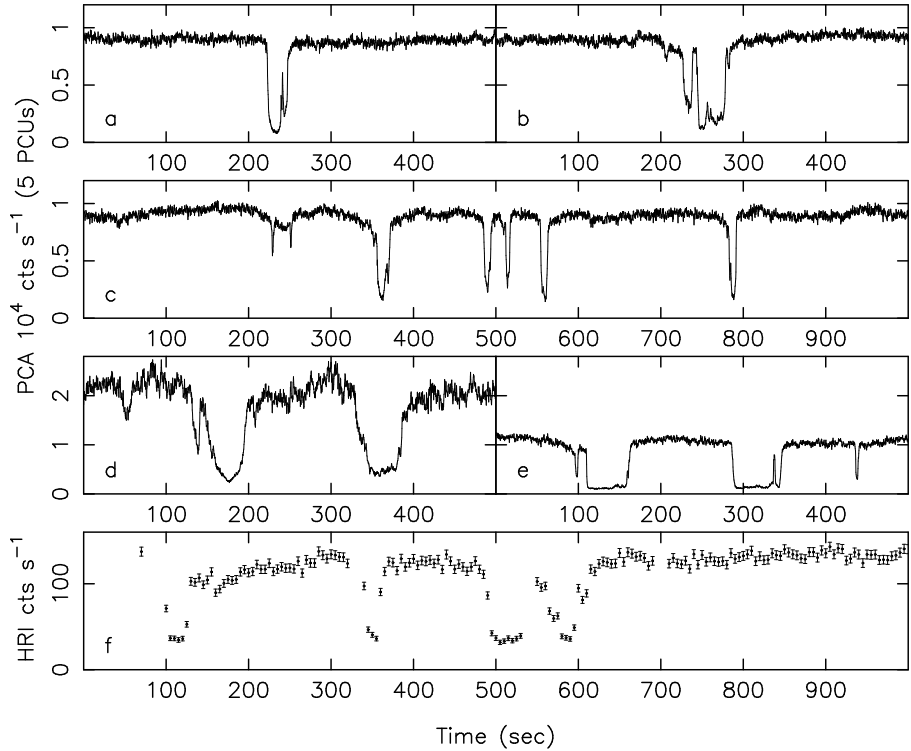}
}
\caption{Collection of light curves from other instruments in which dips have been seen.
a.)--e.): RXTE/PCA; f.): ROSAT/HRI. T=0\,sec corresponds to
UTC 1997 January 5 07:56:19, 08:23:49, 09:46:19, 1996 June 20 13:01:35, 1997 February 26 21:53:35,
and 1997 March 12 03:38:01, for the light curves in a.)--f.), respectively. 
The RXTE/PCA (2--60\,keV) light curves have a time resolution of 0.25\,sec, while the
ROSAT/HRI (0.1--2.4\,keV) light curve has a time resolution of 5\,sec.
Note that the y-axis in d.) and e.) differs from a.)--c.).
}
\end{figure}

In Fig.~4 we show our collection of X-ray light curves from other instruments 
(a.--e.: RXTE/PCA, f.: ROSAT/HRI) during which clear dips were seen. 
Again, the dips are short, i.e.\ they have durations of up to a minute, and deep.
Some of the light curves (e.g.\ Fig.~4c.) show already a slight depression 
in the intensity just before and after the dips.
The main differences in the dip profiles can be discerned by comparing
Figs.~4d.\ and e. Apart from a difference in the out-of-dip intensity,
the dips on 1996 June 20 (Fig.~4d) are more gradual and somewhat less deep (down to $\sim$10\%\
of the out-of-dip intensity), than those that occurred on 1997 January 5 (Fig.~4a--c) 
and February 26 (Fig.~4e).

We note that ASCA observed GRO\,J1655$-$40 from 1997 February 26 00:46:49 to 
February 28 16:50:49, i.e.\ around the time of the RXTE/PCA observations on 1997 February 26. 
However, the X-ray coverage was rather sparse between 
orbital phases 0.76--0.94, and no X-ray dips could be found at these or other phases.

We determined all times of the dips found in the observations described in 
this paper; the time and orbital phase 
ranges of their occurrences are displayed
in Table~2 and indicated in the upper panel of Fig.~1. 
The total orbital phase range between which dips occur is 0.68--0.92.
In Fig.~5 we plot the histogram of the orbital phase occurrences of 
all the dips. The occurrences cluster near $\phi_{\rm orb}\sim 0.8$ and the distribution
is consistent with being Gaussian. 
It may be interesting to note that all dips were only seen during the first halfs of the 
two stages of the outburst, i.e.\ dips occurred during JD\,2450205--331, i.e.\
1996 May 1 to August 17, and JD\,2450453--567, i.e.\ 1997 January 4 to April 28 (Table 2, Fig.~1). 

\begin{table}
\caption{GRO\,J1655$-$40 dip occurrence times}
\begin{tabular}{lccc}
\hline
Time range & $\phi_{\rm orb}$$^a$ range & Cycle$^b$ & Instrument \\
\multicolumn{2}{l}{(JD$-$2\,450\,000)} & \multicolumn{1}{l}{~} & \multicolumn{1}{l}{~} \\
\hline
205.0785--205.0796 & 0.86       & 139 & RXTE/ASM \\
220.5827--220.5850 & 0.77       & 145 & RXTE/ASM \\
223.1122--223.3692 & 0.81--0.83 & 146 & RXTE/ASM \\
225.7172--225.7183 & 0.73       & 147 & RXTE/ASM \\
228.6535--228.6546 & 0.85       & 148 & RXTE/ASM \\
254.8466--255.0469 & 0.84--0.92 & 158 & RXTE/ASM, RXTE/PCA \\
280.7353           & 0.71       & 168 & RXTE/ASM \\
296.7989           & 0.84       & 174 & RXTE/ASM \\
314.9044           & 0.75       & 181 & RXTE/ASM \\
317.6420--317.7279 & 0.79--0.82 & 182 & BSAX/WFC \\
330.9884           & 0.88       & 187 & BSAX/WFC \\
453.8334--453.9162 & 0.74--0.77 & 234 & RXTE/PCA \\
487.9261--487.9909 & 0.74--0.77 & 247 & RXTE/ASM \\
506.4138--506.4159 & 0.80       & 254 & RXTE/PCA \\
519.6519--519.6574 & 0.85       & 259 & ROSAT/HRI \\
524.7309--524.7795 & 0.78--0.80 & 261 & BSAX/WFC \\
553.3015--553.5831 & 0.68--0.79 & 272 & BSAX/WFC, RXTE/ASM \\
556.2916--556.2927 & 0.82       & 273 & RXTE/ASM \\
566.4994           & 0.71       & 277 & RXTE/ASM \\
\hline
\multicolumn{4}{l}{$^a$\,\,Photometric orbital phase (Orosz \&\ Bailyn 1997, van der Hooft et al.\ 1998).} \\
\multicolumn{4}{l}{$^b$\,\,Number of binary cycles since JD~2449838.4198 (van der Hooft et al.\ 1998).} \\
\end{tabular}
\end{table}

\begin{figure}
\centerline{
\psfig{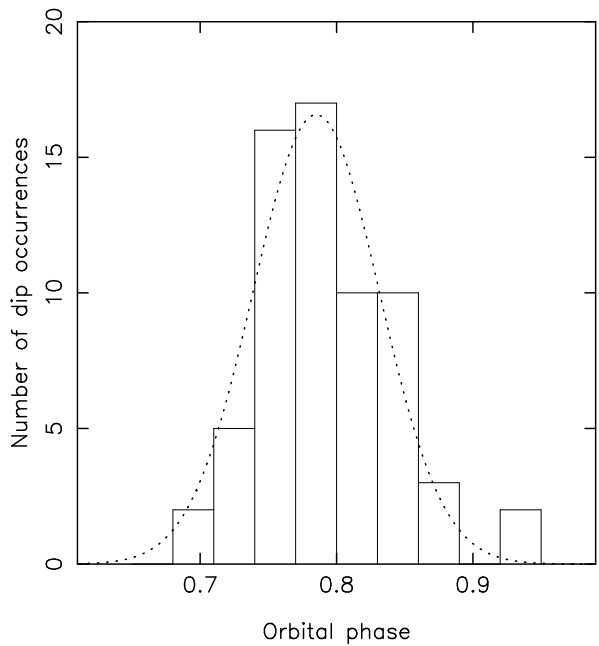}
}
\caption{
Histogram of the orbital phases at which the 65 dips we collected occur (see text).
Drawn also with a dotted line is a Gaussian with a mean of $\mu=0.785$ and standard deviation of
$\sigma$=0.046.
}
\end{figure}

\subsection{From binary orbit to binary orbit}

Due to the long continuous coverage of the WFC it is now possible for the first time 
to follow the behaviour from binary orbit to the next binary orbit.
If one looks at the light curves exactly one or several binary orbit(s) earlier and/or 
later than the binary orbit which did display dips, no dips are found.
This is illustrated in Figs.~2 and 3. In Fig.~2 we show
the WFC light curves from orbital phases 0.65--0.95 taken one binary orbit apart.
Clearly, the dips observed in binary cycle 
182 (left panel) just after orbital phase 0.79 are 
not there in binary cycle 183 (right panel) at the same orbital phase.
In Fig.~3 we show light curves 6 binary
orbits apart from each other. The strong dipping behaviour in binary cycle 272
(right panel) is clearly not repeated at exactly the same phases in binary cycle 266
(left panel). 

The dipping behaviour seen in cycle 272 is stronger, 
i.e.\ much more dips occur, than in cycle 182, as well as compared to other binary orbits
covered by the WFC. We attribute this to the transient nature of the dips, since we do not
think that most or all of the dips are missed by earth occultations
or SAA passages.

\begin{figure}
\centerline{
\psfig{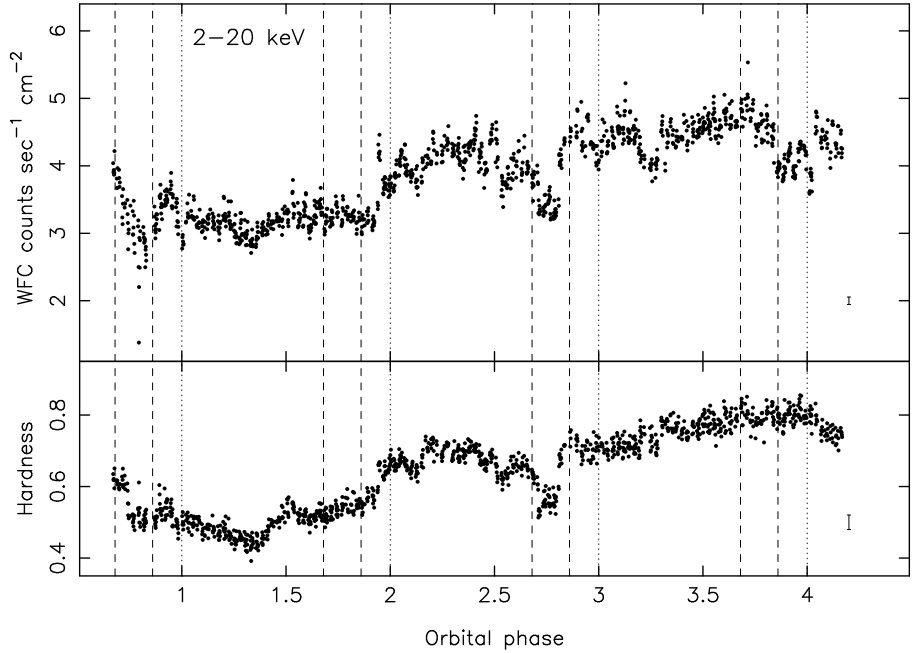}
}
\caption{WFC light curve (upper panel; 2--20\,keV) and hardness curve 
(lower panel; hardness is the ratio of the count rates in the 5--20\,keV and
2--5\,keV bands) of GRO\,J1655$-$40 during the uninterrupted (except for earth occultations 
or South-Atlantic anomoly pasages) nine-day Galactic Center campaign in 1996
as a function of orbital phase. The time resolution in both panels is 5\,min.
The start of the observations corresponds to 1996 Aug 21, 19:46:01.
Two consecutive dashed lines denote the phase ranges between which dipping behaviour has 
been observed. The dotted lines denote zero orbital phase, i.e.\ when the secondary star 
is closest to us, in front of the accretion disk and compact object.
Typical uncertainties are indicated in the lower right corners of the panels.
}
\end{figure}

To illustrate the changes per binary orbit further we constructed
a light curve in the 2--20\,keV band (Fig.~6, upper panel) and hardness curve 
given by the ratio of 5--20\,keV and 2--5\,keV (Fig.~6, lower panel) 
of the first nine consecutive days of observations in 1996.
This is a unique data set from a continuous monitoring campaign of the Galactic Center
(Cornelisse et al.\ 2000, in preparation). 
For clarity we have indicated the phase intervals of observed dipping behaviour
(dashed lines) and the time when the companion star is closest to us, i.e.\ phase zero
(dotted lines). Overall, both the intensity and hardness increased during the nine-day period.
No clear repeatable features per orbit are seen, apart from a depression
in the intensity near orbital phases 0.6--0.9 and 2.6--2.9. Moreover, the dip features which 
occur in the beginning of the
observation (see also Fig.~2), are not repeated in subsequent orbits, 
although some might
have been missed due to earth occultations or SAA passages (see also above).
Note that the depressions in the light curves near orbital phases 0.6--0.9 and 2.6--2.9 are accompanied by a softening in the hardness values. 
On the other hand, no clear hardness changes are discerned 
during the depressions in the light curves near orbital phases 3.3--3.4 and 3.9--4.0.

\section{Discussion}

\subsection{Orbital phase dependence}

Combining all times of X-ray dip occurrences 
during the 1996/1997 outburst of GRO\,J1655$-$40
we find that the dips occur between orbital 
phases 0.68 and 0.92, i.e.\ extending the dip range 
with respect to that quoted by \mbox{Kuulkers} et al.\ (1998a). 
The occurrence of the observed dips concentrate near orbital phases 0.75--0.80.
Similar deep short dips
have now been seen in other 
X-ray transient systems: 4U\,1630$-$47 (Tomsick et al.\ 1998; \mbox{Kuulkers} et al.\ 1998a) and 
Cir X-1 (Shirey et al.\ 1999). As noted by \mbox{Kuulkers} et al.\ (1998a)  
the duration of the dips is also of the same order
as those seen in Cyg\,X-1 and Her\,X-1 (e.g., Kitamoto et al.\ 1984, Leahy 1997), but is shorter
than those typically seen in LMXB dip sources (see e.g.\ Parmar \&\ White 1988).

For the first time, we were able to obtain a nearly-continuous view of the system
for several binary orbits during different parts of the outburst.
On one occasion we found long irregular dipping behaviour 
for a period of $\sim$9.2\,hr, which spans about $\sim$0.15 in orbital phase.
This duty cycle is much larger than quoted by Tomsick et al.\ (1998)
for GRO\,J1655$-$40; it is similar to that seen in low-mass X-ray binary dip sources 
(`dippers'; e.g.\ Parmar \&\ White 1988;
White et al.\ 1995).

Do the dips recur every binary orbit? At certain times they do, as was shown
by \mbox{Kuulkers} et al.\ (1998b) to be the case during the early part of the 1996/1997 outburst
using the RXTE/ASM light curves. 
Our collection of dip times suggests the same during other parts of the outburst 
(see Table 2). If one compares in detail the light curves from binary orbit to binary orbit, it is 
clear that the dips do not persist at exactly the same phase. Moreover, if the
strong dipping behaviour seen in binary cycle 272 were present every orbit, we would certainly
have seen this in the WFC light curves. In contrast most of the WFC light curves
do not show evidence for any strong dipping behaviour.
Such a transient nature of the dip activity
has been noted before as well in the other LMXB dip sources, see
e.g.\ Parmar et al.\ (1986) for the case of EXO\,0748$-$676. 
We note that it looks as if dipping only occurred during the first halfs of the 
two stages of the outburst, i.e.\ they appear during the rise and plateau phases of the
outburst. 

\subsection{Stream-disk interaction}

\begin{figure}
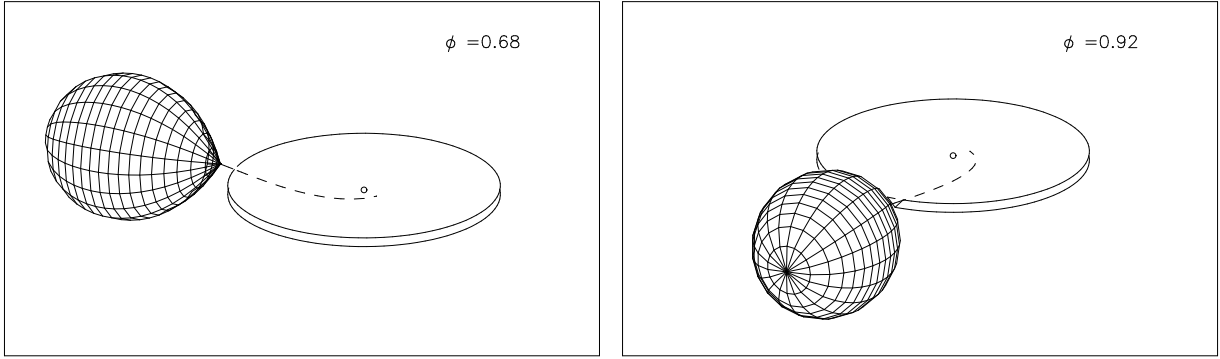

\centerline{
\hspace{2.5cm}
\psfig{figure=1655_wfc_fig7a.ps,bbllx=78pt,bblly=258pt,bburx=534pt,bbury=534pt,width=8cm,clip=yes}
\hspace{.0cm}
\psfig{figure=1655_wfc_fig7b.ps,bbllx=78pt,bblly=258pt,bburx=534pt,bbury=534pt,width=8cm,clip=yes}
}
\caption{View of GRO\,J1655$-$40 at the extreme ends of the dip phase range, i.e.\
$\phi_{\rm orb}\sim 0.68$ (left panel) and $\phi_{\rm orb}\sim 0.92$ (right panel).
We set the following binary parameters: $q$=0.354 (Shahbaz et al.\ 1999) and
$i$=67.2$^{\circ}$ (van der Hooft et al.\ 1998). We arbitrarily choose a
disk opening angle of 2$^{\circ}$
(see Phillips et al.\ 1999) and a disk radius of 0.85\,R$_{\rm L1}$ (see Orosz \&\ Bailyn
1997). The dashed line shows the ballistic path of a stream from the secondary to the compact object
in the absence of an accretion disk. 
}
\end{figure}

The orbital phase dependence of the dip occurrences 
suggests the dip mechanism to be fixed in the binary orbit.
For illustrative purposes (Fig.~7) we show views 
of the system at the extremes of the dip phase range.
The most straightforward site is the interaction region between the mass transfer
stream from the companion star and the accretion disk (e.g.\ Frank et al.\ 1987).
However, it is not the bulge on the accretion disk itself which obscures the X-ray emitting
region. This would result in a rather long X-ray dip as seen in e.g.\ accretion disk
corona sources (see e.g.\ White et al.\ 1995); moreover, the inclination at which we view 
GRO\,J1655$-$40 is too low for such an effect to be seen (see Frank et al.\ 1987).
Rather the short duration of the dips points to an absorbing medium which is
filamentary in nature and probably situated above and/or below the impact region
(\mbox{Kuulkers} et al.\ 1998b).
A similar filamentary nature has been proposed for the medium in Cyg\,X-1 
that causes the second peak in the distribution of (short) dips with orbital 
phase at $\phi_{\rm orb}$$\sim$0.6 (where phase zero corresponds to superior 
conjunction of the black hole). This peak has been attributed also to a
mass transfer stream from the O-supergiant interacting with the accretion disk
(Ba\l uci\'nska-Church et al.\ 2000). Note that the main peak occurs 
at $\phi_{\rm orb}$$\sim$0.95, which is due to absorption in the wind of the supergiant.

The dynamics of mass transfer streams from the companion up to the place of impact 
onto the accretion disk was first investigated by Lubow \&\ Shu (1975, 1976).
More recently it has become possible to study in more detail the impact
region itself and what happens with the matter after it reaches this region
(e.g.\ Armitage \&\ Livio 1998, and references therein).
The hydrodynamic simulations by 
Armitage \&\ Livio (1998) show that the fate of the stream after it impacts the disk
depends on the efficiency of cooling in the shock-heated gas created by the impact.
If the cooling is efficient, the upper and/or lower parts of the stream are able to freely
overflow the disk, unless the disk rim is too thick. 
This ballistic stream will reimpact the disk near to its closest 
approach to the compact object (see also Frank et al.\ 1987, Lubow 1989).
Observationally this may lead to rather broad dips 
occurring near orbital phase 0.6 (Lubow 1989, see also Armitage \&\ Livio 1996).
If cooling is inefficient than the stream `splashes' onto the disk rim, i.e.\
the material overflowing the disk is a coherent stream but rather
erratic. Moreover, more material is able to reach greater heights from the disk,
which leads to substantially higher absorption columns with respect to the former case, 
for lines-of-sight well away from the disk plane. At such lines-of-sight one expects
a plethora of dips to be observed from orbital phases $\sim$0.7--1.0, with a peak near orbital
phase 0.8 (Armitage \&\ Livio 1996, see also Frank et al.\ 1987). Moreover, in the latter case
one expects a bulge extending along the disk rim as well (Armitage \&\ Livio 1998).

The dips observed during the 1996/1997 outburst of GRO\,J1655$-$40 all share the characteristics
of the case of non-efficient cooling: they are short, show absorption columns up to
$\sim$10$^{24}$~atoms\,cm$^{-2}$ (\mbox{Kuulkers} et al.\ 1998a) and the distribution of
dip occurrence times extends from binary orbital phases $\sim$0.7--0.9, with a peak near 
binary orbital 0.8.

Apart from the `splashing' material causing the deep short dips, 
the presence of an extended bulge along the disk rim is supported by the observed
decrease in optical polarisation near orbital phase
$\sim$0.7--0.8 (Gliozzi et al.\ 1998). 
The observations were obtained just after the peak of the second part of the outburst
during 1997 July 2--9. 
The polarisation region must be rather extended and is located primarily
around the inner disk regions.
A thickening of the disk rim or the remnant stream passing under
and over the disk then obscures smoothly this polarisation region
(Gliozzi et al.\ 1998).

ASCA observed a $\sim$4\,hr drop in intensity 
by $\sim$75\%\ in the light curve of GRO\,J1655$-$40 on 1994 August 23 (Ueda et al.\ 1998).
These observations were obtained just after
the peak of the first hard X-ray and radio outburst (Harmon et al.\ 1995;
Tavani et al.\ 1996).
Using the ephemeris by van der Hooft et al.\ (1998), the ASCA 
dip occurred between orbital phases $\sim$0.53--0.61, i.e.\ outside the 
phase range we see for the short and deep dips during the 1996/1997 
outburst\footnote{We note that an eclipse-like feature in the optical light
curves was found on 1994, August 17 by Bailyn et al.\ (1995a), i.e.\ 
close in time to the ASCA dip; a possibly similar optical feature occurred 
one binary orbit earlier (Bianchini et al.\ 1997).
The eclipse-like light curve of Bailyn et al.\ (1995a) lasted from 
orbital phase $\sim$0.0--0.1. 
Bianchini et al.\ (1997) suggested this to be due to a bright spot region, 
or an extended optically thick disk rim shielding part of the disk.}. 
No deep short dips have been reported for the outbursts which occurred in 1994/1995, however,
the soft X-ray ($<$10\,keV) coverage during these outbursts was very sparse. 
The long ASCA dip is in agreement with what is 
expected when the stream encounters a cold
accretion disk rim, so that part of the stream can continue to flow
above and under the accretion disk (see above). 
According to Ueda et al.\ (1998), the material causing the dip
by absorption must consist of slightly ionized (warm) and non-ionized (cold) 
material, which is also in line with the fact that the stream will re-impact the 
disk sufficiently close to the region where ionisation becomes important (Frank et al.\ 1987). 

We suggest that the difference between the observed X-ray dips during the 1994 and 1996/1997 
outbursts may be due to the efficiency in irradiating the outer disk regions. 
In 1994 the disk rim may have had a shape such that it does not see the X-ray emitting regions
(i.e.\ self-screening by the disc of its outer regions; see Dubus et al.\ 1999),
possibly due to a large enhancement of mass transfer from the secondary, and the 
gas at the stream impact region could cool efficiently. During the 1996/1997
outburst the outer disk regions were less affected by the mass transfer stream from
the secondary, which were therefore more easily prone to X-ray irradiation by the X-ray emitting
region, keeping the outer disk hot.

\begin{acknowledgements}

EK thanks Jerry Orosz and Frank Verbunt for comments on an earlier
draft of this paper and 
Keith Horne for supplying the source code of `The CV Eclipse Movie
(not showing at a theatre near you)'. AKH is supported by a Hong Kong
Oxford Scholarship.
The BeppoSAX satellite is a joint Italian and Dutch programme.

\end{acknowledgements}

\end{document}